# Soliton collisions in soft magnetic nanotube with uniaxial anisotropy

N A Usov[1,2]

[1]National University of Science and Technology «MISIS», 119049, Moscow, Russia
[2]Pushkov Institute of Terrestrial Magnetism, Ionosphere and Radio Wave Propagation, Russian Academy of Sciences,
(IZMIRAN) 142190, Troitsk, Moscow, Russia

**Abstract**

The structure of stable magnetic solitons of various orders in soft magnetic nanotube with uniaxial magnetic anisotropy has been studied using numerical simulation. Solitons of even order are immobile in axially applied magnetic field. Odd solitons show decreased mobility with respect to that of head-to head domain wall. Solitons of various orders can participate in nanotube magnetization reversal process. Various coalescence and decomposition processes in soliton assembly are considered. It is shown that the general magnetization state of magnetic nanotube consists of chains of magnetic solitons of various orders.

## 1. Introduction

Magnetic nanowires and nanotubes attract considerable research interest due to their fundamental magnetic properties and potential application in various areas of nanotechnology, for example, for production of advanced spintronic and sensor devices[1], as well as for high density magnetic recording[2]. An array of magnetic nanotubes can be fabricated chemically[3], or by atomic layer deposition of various ferromagnetic materials in porous anodic aluminum oxide[4-6]. An isolated magnetic nanotube can also be created by coating a carbon nanotube grown at the end of a magnetic force microscopy cantilever with a magnetic material[7,8]. Using carbon nanotube tips coated with a ferromagnetic metal one can substantially improve a lateral resolution in the magnetic force microscopy observations[9,10].

The magnetization distribution and domain wall (DW) dynamics in magnetic nanotubes have recently been studied both theoretically[11-19] and experimentally[20-23]. The control of the magnetic domain state and magnetization reversal process in magnetic nanotubes is especially important for various applications. Using the numerical simulation, in the present paper it is shown that in a magnetic nanotube with uniaxial anisotropy there may exist magnetic solitons of various orders alongside with usual DWs. The magnetic solitons can participate in magnetization reversal process. Various coalescence and decomposition processes in the soliton assembly are considered. It is shown that in general magnetization state of magnetic nanotube consists of chains of magnetic solitons of various orders.

## 2. Numerical simulation

In the lowest energy state a nanotube with positive uniaxial anisotropy constant, $K > 0$, is uniformly magnetized along its length except for the curling states[12] arising near the tube ends due to the influence of a strong demagnetizing field. These curling states initiate nucleation[15] of the head-to-head DW at the tube end in external magnetic field $H_z$ applied opposite to initial uniform tube magnetization. If $H_z > H_n$, where $H_n$ is the DW nucleation field, the DW starts at the tube end and propagates along the tube to accomplish magnetization reversal process.

The unit magnetization vector of the head-to-head DW in a thin soft magnetic nanotube was shown [12] to be approximately described in the cylindrical coordinates $(r, \varphi, z)$ as

$$\alpha_z = \cos\theta(z) = \tanh(z/\delta);$$
$$\alpha_\varphi = \pm\sqrt{1-\alpha_z^2}\;; \qquad \alpha_r \approx 0, \qquad (1)$$

where $\delta$ is the DW width. It is easy to see that the head-to-head DW is charged. It resembles usual Neel type DW in a thin magnetic film, since the unit magnetization vector rotates mostly within the cylindrical surface. On the other hand, the head-to-head DW is a 180° isolated magnetic soliton (MS), as the angle $\theta(z)$ changes from 0 to 180° when $z$ coordinate passes through the DW center. The fact that both signs of the $\alpha_\varphi$ component in Eq. (1) are possible enables one to construct topologically stable MSs of higher order, $N > 1$. For MS of the $N$-th order the unit magnetization vector rotates by 180$N$ degrees as a function of $z$ coordinate. One has $N = 1$ for the usual DW, Eq. (1), $N = 2$ corresponds to 360° MS, $N = 3$ gives 540° MS, etc.

In this paper the structure and properties of the high order MSs in thin magnetic nanotube with uniaxial magnetic anisotropy have been studied by means of 2D numerical simulation using Landau – Lifshitz – Gilbert (LLG) equation[12]. The magnetic parameters of the nanotube are those of a soft magnetic material, i.e. saturation magnetization $M_s = 1000$ emu/cm$^3$, uniaxial anisotropy constant $K = 10^5$ erg/cm$^3$, and exchange constant $A = 10^{-6}$ erg/cm. The outer tube diameters vary in the range $D = 60 - 120$ nm, the tube thickness being $\Delta D = 8 - 20$ nm. The tube length equals $L = 2400 – 4000$ nm. It can be shown that the further increasing of the tube length at a fixed diameter does not change the numerical simulation results. Therefore, the results obtained correspond actually to very long nanotubes with aspect ratio $L/D \gg 1$.

Due to cylindrical symmetry of the magnetization distribution the unit magnetization vector of the nanotube does not depend on the azimuthal angle $\varphi$. Therefore, one can perform 2D numerical simulations subdividing a long piece of a nanotube into an array $N_r \times N_z$ of toroidal numerical cells. The size of the toroidal numerical cell is given by $dz = dr = 2$ nm, i.e. much smaller than the exchange length, $L_{ex} = \sqrt{A/M_s} = 10$ nm. Magnetostatic interactions as well as self-energies of the toroidal elements are calculated preliminarily for a given array in a manner similar to the usual case of small cubic elements. It enables one to calculate the demagnetizing field for any given magnetization distribution among the toroidal elements. In the absence of external magnetic field the final magnetization distribution is supposed to be stable under the condition

$$\max_{(1\leq i \leq N)}\left|[\vec{\alpha}_i, \vec{H}_{ef,i}]\right| < 10^{-8}; \quad i = 1, 2, \ldots N,$$

where $N = N_r N_z$ is the total number of the numerical cells in the nanotube. This means that the unit magnetization vector $\boldsymbol{\alpha}$ in each numerical cell is parallel with a sufficiently high accuracy to the effective magnetic field $H_{ef}$ in the same numerical cell. For dynamic micromagnetic simulations the magnetic damping constant in the LLG equation is chosen to be $\kappa = 0.1$.

## 3. Results and discussion

The micromagnetic structure of MS of $N$-th order can be approximated by means of the Ritz function, $\cos(\theta(z)/N) = \tanh(z/\delta)$. For example, for 360° MS ($N = 2$) the components of the unit magnetization vector are given by

$$\alpha_r = 0; \quad \alpha_\varphi = \sin\theta = \frac{2\sinh(z/\delta)}{\cosh^2(z/\delta)}$$
$$\alpha_z = \cos\theta = \frac{\sinh^2(z/\delta)-1}{\cosh^2(z/\delta)}. \quad (2)$$

For 540° MS ($N = 3$) one has the following relations

$$\alpha_r = 0; \quad \alpha_\varphi = \frac{3\sinh^2(z/\delta)-1}{\cosh^3(z/\delta)};$$
$$\alpha_z = \frac{\sinh^3(z/\delta)-3\sinh(z/\delta)}{\cosh^3(z/\delta)}. \quad (3)$$

Similar formulas can be easily obtained for MS of a higher order.

The magnetization distributions, Eq. (2), (3), can be used as initial magnetization states to calculate actual micromagnetic structures of 360° and 540º MSs. Fig. 1a shows the micromagnetic structure of 360° MS obtained numerically in the magnetic nanotube with $D = 80$ nm and $\Delta D = 10$ nm in comparison with the Ritz function given by Eq. (2). The micromagnetic structure of the 540° MS in the nanotube with $D = 120$ nm, $\Delta D = 8$ nm is shown in Fig. 1b in comparison with the corresponding Ritz function, Eq. (3). One can see that the Ritz functions assumed are in a satisfactory agreement with the 2D numerical simulations performed. These Ritz functions are appropriate as they take into account the topological structure of the high order solitons. Note also that for both MSs shown in Fig. 1 the radial component of the unit magnetization vector calculated numerically is close to zero.

It is found that for a thin nanotube with thickness $\Delta D \ll D$, the characteristic MS width $\delta$ is only slowly dependent on the outer tube diameter $D$. For example, for the case of 540º MS the parameter $\delta$ decreases from $\delta = 128$ nm to $\delta = 101$ nm, when the outer tube diameter increases from $D = 80$ nm to $D = 120$ nm. At the same time, the characteristic width of 360º MS remains nearly constant, $\delta \approx 40$ nm, in the above range of the outer tube diameters.

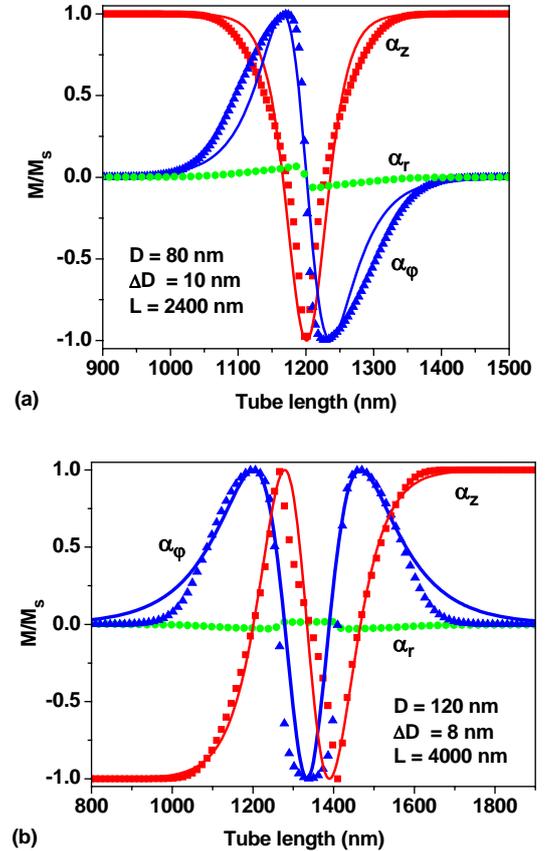

Figure 1. (a) Micromagnetic structure of 360º MS located within a long magnetic nanotube; dots are the result of numerical simulation, solid lines are drawn according to equation (2), the characteristic MS width being $\delta = 40$ nm. (b) The same for 540º MS; the characteristic MS width equals $\delta = 101$ nm.

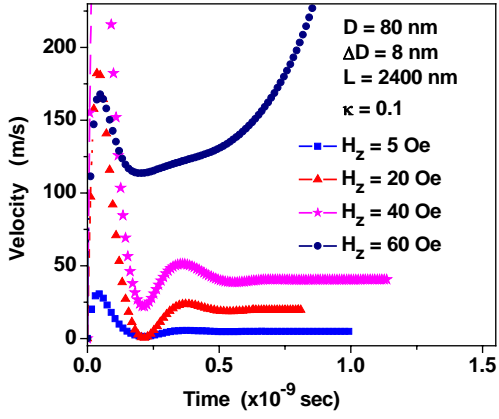

Figure 2. Velocity of 540º MS at various values of applied magnetic field.

The 360° MS is immobile in external magnetic field parallel to the tube axis, because it seperates the domains with the same sign of the $\alpha_z$ component. For the same reason, all even solitons, $N = 2k$, are immobile in external magnetic field. Whereas, the odd solitons, $N = 2k+1$, have finite mobility in applied magnetic field. Fig. 2 shows the velocity of the 540º MS as a function of time for different values of the applied magnetic field. The average speed of the soliton is calculated as the time derivative of the $\alpha_z$ component averaged over the total length of the nanotube, $v_z = d<\alpha_z>/dt$. As can be seen in Fig. 2, in the range of moderate magnetic fields, $H_z \leq 40$ Oe, after a short initial period of time the velocity of the 540º MS approaches a stationary value. The stationary MS velocity is proportional to the applied magnetic field. However, the mobility of the 540º MS in the nanotube with $D = 80$ nm and $\Delta D = 8$ nm is found to be small, $\mu \approx 1.0$ m/s/Oe. This value is almost an order of magnitude less than the mobility of 180º MS ($\mu \approx 10.0$ m/s/Oe) in the same nanotube and at the same value of the damping parameter[12]. It is found also that the steady movement of 540º MS is impossible if external magnetic field exceeds some critical value, $H_z \geq H_c \approx 50$ Oe. In this case the 540º MS decomposes into immobile 360º MS and fast 180º MS. As shown in Fig. 2, in sufficiently high magnetic field, $H_z = 60$ Oe, for a typical time interval of the order of $10^{-9}$ s the velocity of the 540º MS does not approach a stationary value, but continues to increase because of the acceleration of the detached 180º MS.

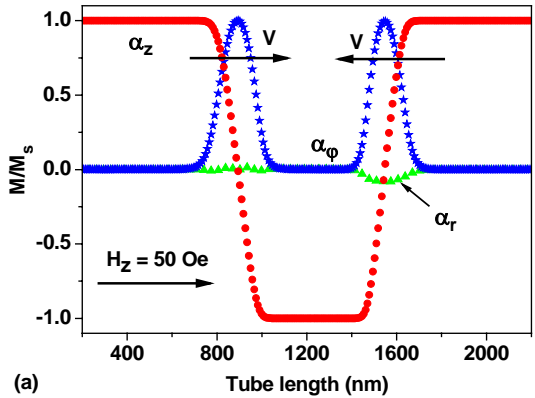

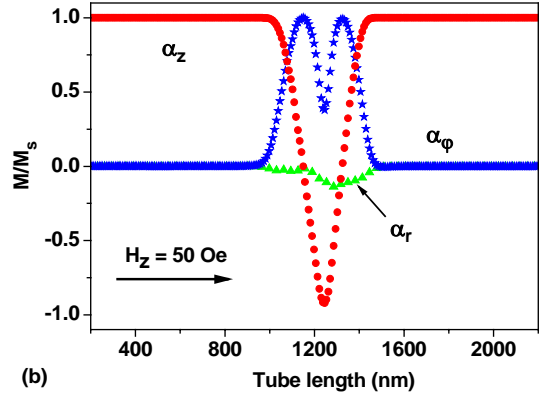

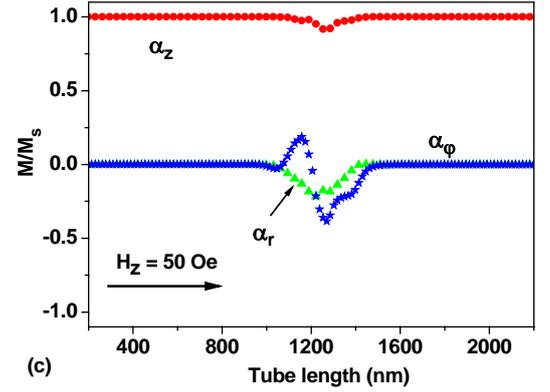

Figure 3. Annihilation of head to head and tail to tail domain walls with the same sign of the $\alpha_\varphi$ component in the nanotube with sizes $60 \times 80 \times 2400$ nm in applied magnetic field $H_z = 50$ Oe.

Interestingly, the collision of two 180º MS in an external magnetic field occurs in different ways, depending on the type of the interacting solitons. As Fig. 3 shows, if the 180º MSs have the same sign of the $\alpha_\varphi$ component, the collision of these MSs in the applied magnetic field leads to their annihilation. Actually, for magnetic field direction shown in Fig. 3a, the magnetic domain with $\alpha_z = -1$ is energetically unfavorable. Therefore, in applied magnetic field $H_z = 50$ Oe the head to head MS moves to the right, whereas tail to tail MS moves to the left, as the panels (a), (b) of Fig. 3 show. Due to the MS annihilation the ground uniform magnetization state, $\alpha_z = 1$, in the main part of the nanotube appears (see panel (c) in Fig. 3). On the other hand, if 180º MSs have opposite signs of the $\alpha_\varphi$ component, the immobile 360º MS appears as a result of their collision. This process is shown in successive panels (a) – (c) of Fig. 4.

It was found that for the nanotubes studied the immobile 360º MS is stable only in the range of magnetic fields $|H_z| < 100$ Oe. In reversed magnetic field $H_z = -120$ Oe, opposite to the tube magnetization outside the soliton, 360º MS decomposes into two fast 180º MSs that propagate to opposite ends of the nanotube.

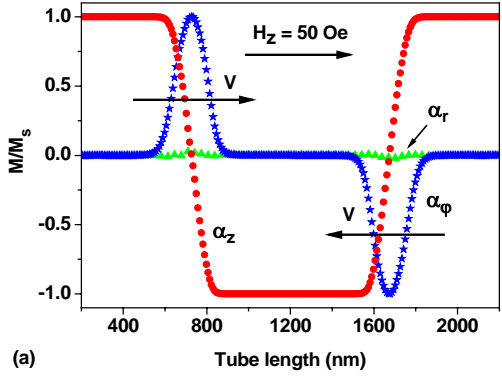

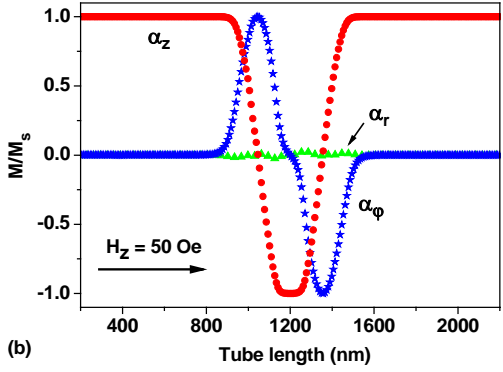

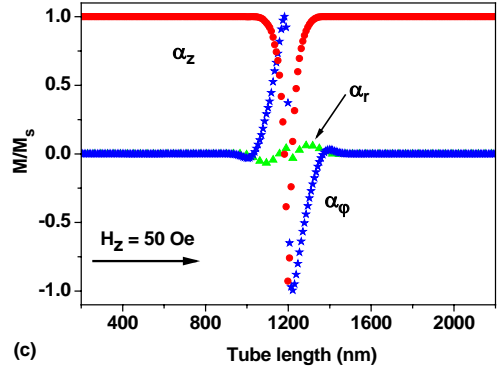

Figure 4. Creation of the 360º immobile MS ($N = 2$) due to collision of the head to head and tail to tail domain walls in magnetic nanotube in applied magnetic field $H_z = 50$ Oe. Contrary to figure 3, these MSs have opposite signs of the $\alpha_\varphi$ component.

More complicated process is shown in Fig. 5, where mobile 180° MS approaches immobile 360° MS in applied magnetic field $H_z = 50$ Oe. The dynamic calculations are carried out in a long nanotube with outer diameter $D = 80$ nm and thickness $\Delta D = 10$ nm. If the sign of the $\alpha_\varphi$ component of 180° MS is positive, as shown in Fig. 5a, the interaction of 180° and 360° MSs leads to creation of 540° MS. But at $H_z = 50$ Oe that is higher then the critical magnetic field for stationary movement of the 540° MS, the latter becomes unstable and transforms into immobile 360° MS and fast 180° MS, as Fig. 5b shows. The 180° MS propagates finally through the nanotube to complete the magnetization reversal process. On the other hand, if $\alpha_\varphi < 0$ the result of collision of mobile 180° MS and immobile 360° MS is a single 180° MS that propagates to the right end of the nanotube.

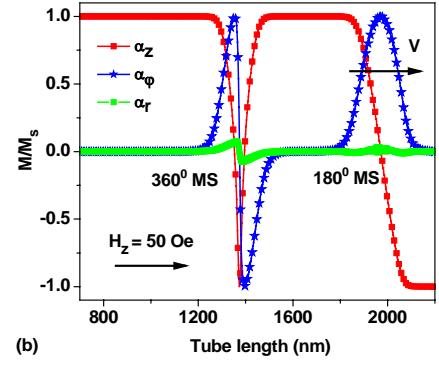

Figure 5. Collision of the mobile 180° MS and immobile 360° MS in applied magnetic field $H_z = 50$ Oe: (a) before soliton interaction; (b) after interaction.

Evidently, 1D magnetic solitons can be considered as excitations of a vacuum state. For a uniformly magnetized nanotube with uniaxial anisotropy there are two different vacuum states, i.e. $\alpha_z = +1$ and $\alpha_z = -1$. For definiteness, let us consider the vacuum state $\alpha_z = +1$. The simplest excitations of this vacuum state are different pairs of 180° MS. Schematically they can be represented in the following manner

$(\alpha_z = +1|\alpha_\varphi = +1|\alpha_z = -1) + (\alpha_z = -1|\alpha_\varphi = +1|\alpha_z = +1)$; (4a)

$(\alpha_z = +1|\alpha_\varphi = +1|\alpha_z = -1) + (\alpha_z = -1|\alpha_\varphi = -1|\alpha_z = +1)$; (4b)

$(\alpha_z = +1|\alpha_\varphi = -1|\alpha_z = -1) + (\alpha_z = -1|\alpha_\varphi = +1|\alpha_z = +1)$; (4c)

$(\alpha_z = +1|\alpha_\varphi = -1|\alpha_z = -1) + (\alpha_z = -1|\alpha_\varphi = -1|\alpha_z = +1)$. (4d)

Taking into account that the density of a volume magnetic charge is given by $q = -M_s div \vec{\alpha}$, it is easy to prove that the head-to-head solitons $A = (\alpha_z = +1|\alpha_\varphi = +1|\alpha_z = -1)$ and $B = (\alpha_z = +1|\alpha_\varphi = -1|\alpha_z = -1)$ possess positive total magnetic charge, $Q = 2M_sS$, where $S$ is the area of the nanotube cross-section. In the positive magnetic field $H_z > 0$ they move from the left to the right. On the other hand, tail-to-tail solitons $C = (\alpha_z = -1|\alpha_\varphi = +1|\alpha_z = +1)$ and $D = (\alpha_z = -1|\alpha_\varphi = -1|\alpha_z = +1)$ possess negative total magnetic charge, $Q = -2M_sS$. Thus, in the positive magnetic field they move in the opposite direction, i.e. from the right to the left. Evidently, all four 180° MS have different micromagnetic structures, as solitons $A$ and $C$ have positive sign of the $\alpha_\varphi$ component, whereas for solitons $B$ and $D$ the sign of $\alpha_\varphi$ component is negative. It is interesting to note that the properties of solitons $A$ and $C$ resemble those of the positron and the electron, respectively. Indeed, they have total magnetic charges of the opposite sign and their collision leads to annihilation, as Fig. 3 shows. Similarly, the properties of solitons $B$ and $D$ resemble those of the proton and the anti-proton, correspondingly. They also have magnetic charges of the opposite sign and annihilate in collision process too. Furthermore, the collision of solitons $A$ and $D$ leads to creation of stable 360° MS with zero total magnetic charge (see Fig. 4). In the same terminology, the latter soliton corresponds to an "anti-hydrogen" atom. However, the collision of solitons $B$ and $C$ creates stable 360° MS that is an analog of a hydrogen atom in the 1D Universe of magnetic solitons.

Fig. 6a shows schematically 4 different types of 180° MS existing in a magnetic nanotube with uniaxial anisotropy. Fig. 6b summarizes the basic collision processes in an assembly of 1D magnetic solitons. As Fig. 6c shows, a proper collection of the 360° MSs can be termed as a stable "magnetic solid".

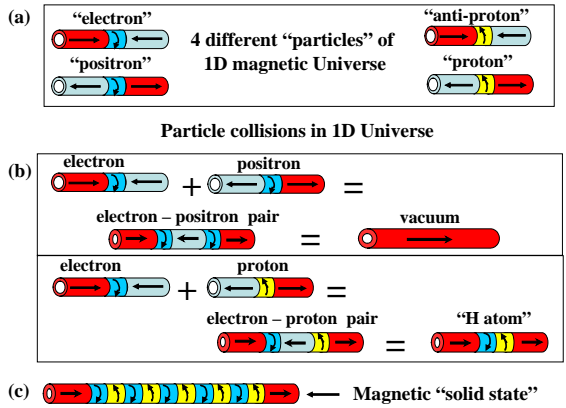

Fig 6. 1D Universe of magnetic solitons existing in a soft magnetic nanotube with uniaxial anisotropy.

Various stable micromagnetic domain structures can be created in a magnetic nanotube by means of time evolution of random initial magnetization states according to LLG equation in zero magnetic field. An example of such a complicated magnetization distribution is shown in Fig. 7. As Fig. 7 shows, such a domain structure is a chain of magnetic solitons of various orders. One notes the presence of several immobile 360° MSs that can be decomposed only in high enough applied magnetic field.

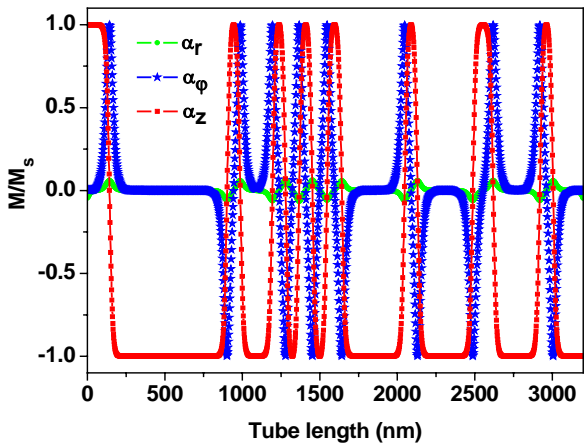

Figure 7. Domain structure consisting of magnetic solitons of various orders in magnetic nanotube with sizes 60 × 80 × 3200 nm.

## 4. Conclusions

In conclusion, in the present paper it is shown that stable MSs of high orders ($N \geq 2$) can exist in soft magnetic nanotube with uniaxial magnetic anisotropy. MSs of even order, $N = 2k$, are immobile in axially applied magnetic field. Odd MSs, $N = 2k+1$, show decreased mobility in applied magnetic field with respect to that of head-to-head DW ($N = 1$). MSs of various orders can participate in nanotube magnetization reversal process. It is found also that for a long nanotube metastable periodic and aperiodic domain structures can be constructed in zero magnetic field. They consist of chains of MS of various orders. It can be shown that MS of similar structure can exist also in magnetic nanowires with circular cross-section. MS of various orders can survive near the defects in magnetic nanotube. It seems possible to create high order MS experimentally through the magneto-static interaction of the tube magnetization with a sharp tip[10] of magnetic force microscope.

### Acknowledgments

The author wishes to acknowledge the financial support of the Ministry of Education and Science of the Russian Federation in the framework of Increase Competitiveness Program of NUST «MISIS», contract № K2-2015-018.

**Figure captions**

Fig. 1. (a) Micromagnetic structure of 360º MS located within a long magnetic nanotube; dots are the result of numerical simulation, solid lines are drawn according to Eq. (2), the characteristic MS width being $\delta = 40$ nm. (b) The same for 540º MS; the characteristic MS width equals $\delta = 101$ nm.

Fig. 2. Velocity of 540º MS as a function of time at various values of applied magnetic field in nanotube with $D = 80$ nm, $\Delta D = 8$ nm and $L = 2400$ nm. Inset shows the stationary velocities of 180º and 540º MSs as the functions of applied magnetic field.

Fig. 3. Annihilation of head to head and tail to tail domain walls with the same sign of the $\alpha_\varphi$ component in the nanotube with sizes $60 \times 80 \times 2400$ nm in applied magnetic field $H_z = 50$ Oe.

Fig. 4. Creation of the 360º immobile MS ($N = 2$) due to collision of the head to head and tail to tail domain walls in magnetic nanotube in applied magnetic field $H_z = 50$ Oe. Contrary to Fig. 3, these MSs have opposite signs of the $\alpha_\varphi$ component.

Fig. 5. Collision of the mobile 180° MS and immobile 360° MS in applied magnetic field $H_z = 50$ Oe: (a) before soliton interaction; (b) after interaction.

Fig. 6. 1D Universe of magnetic solitons existing in a soft magnetic nanotube with uniaxial anisotropy.

Fig. 7. Domain structure consisting of magnetic solitons of various orders in magnetic nanotube with sizes $60 \times 80 \times 3200$ nm.